\documentstyle[12pt]{article}
\begin{document}

\title{A Simple Model for Deep Bed Filtration}
\author{Jysoo Lee and Joel Koplik \\ 
Benjamin Levich Institute and Department of Physics \\
City College of the City University of New York \\
New York, NY 10031}
\date{February 1, 1996}
\maketitle

\begin{abstract}

We present a simple model for deep bed filtration, where particles
suspended in a fluid are trapped while passing through a porous
filter.  A steady state of the model is reached when filter can not
trap additional particles.  We find the model has two qualitatively
different steady states depending on the fraction of traps, and the
steady states can be described by directed percolation.  We study in
detail the evolution of the distribution of trapped particles, as the
number of trapped particles increases.  To understand the evolution,
we formulate a mean field equation for the model, whose numerical
solution is consistent with the behavior of the model.  We find the
trapped particle distribution is insensitive to details of the
formulation of the model.

\vspace{20pt}
\end{abstract}
\newpage

\section{Introduction}
\label{sec:intro}

Deep bed filtration is a well-established process used to separate
solid particles suspended in a fluid \cite{hlg70,g77,dbl88,t89}.  A
dilute suspension is injected into filter made of porous material.
Particles, while flowing through the filter, are trapped inside by
various mechanisms.  The trapped particles can later be recovered by
``cleaning'' the filter.

The quantities of main interests are the filter efficiency (the
fraction of injected particles trapped in filter) and the pressure
drop across filter in order to maintain a constant fluid flow.  As
more particles are trapped in filter, the filter efficiency usually
decreases, and the pressure drop usually increases.  The theory of
deep bed filtration should predict these quantities in terms of
parameters of the system.  In order to build such a theory, one needs
certain knowledge about the dynamics inside filter, e.g., the spatial
distribution of trapped particles.  Unfortunately, such information
has been very limited \cite{lhzp89,h90}.

Recently, Ghidaglia {\em et al} carried out a series of experiments on
deep bed filtration \cite{ggo91,gahg96,gahg96b}.  Instead of a
conventional porous material (e.g., sandstone), they used a random
packing of glass spheres as the filter medium.  The transparency of
glass and the index matched fluid used for the suspension allow direct
visual observation inside the whole filter.  The movements of
particles can be followed in great details.  The setup can be used to
gain valuable information inside the filter, such as the interaction
of particles with porous medium, and the distribution of trapped
particles.  It also becomes a challenge to understand these newly
available quantities.

Network models can be used to study the behavior of particles and a
fluid in a filter \cite{ldg84,rf88,si91,hsd93}.  In a network model,
the inner structure of filter is modeled by pores interconnected by
narrow channels.  A microscopic pressure-flow relation, e.g., Darcy's
law, is assumed across each channel.  Such relations and the external
boundary conditions provide equations for the flow field, which can be
solved numerically.  The motion of a particle is determined by both
the local flow field and the interaction between particles.  The main
advantage of a network model is that it is a good approximation to a
real system.  For example, the flow field and the movement of a
particle can be calculated from microscopic equations with a
reasonable geometry.  The disadvantage, however, is large
computational efforts necessary even for the simulations of a moderate
size system.

Instead, we propose a cellular automata model for deep bed filtration.
The main advantage of the model is, due to its simplicity, that one
can study detailed behaviors of the systems of fairly large size.
Only geometric properties can be obtained from the model.  Also, the
rules for the movement of particles are too simple to capture the
detailed interactions of real particles.  For example, the actual flow
field in filter constantly changes with the movement of particles.
Such changes are mostly ignored in the present model.  We thus expect
that only those aspects of the behavior of the model which are not
sensitive to the details of the rules can be compared with
experiments.  Such comparisons are necessary to establish the validity
and the limitations of the model.

As the number of trapped particles increases, the model reaches a
steady state in which no additional particle can be trapped.  The
steady state can, as we shall see, be described by directed
percolation (DP).  The qualitative behavior of the steady state is
different depending on a parameter $p$, the fraction of trapping
bonds.  If $p$ is less than threshold $p_c$, a newly injected particle
simply pass through the filter without being trapped.  On the other
hand, if $p > p_c$, all the paths leading to an exit are blocked.  We
study in detail the evolution of the distribution of trapped particles
for various values of $p$.  To understand the behavior of the model,
we construct a mean field differential equation for the evolution of
the distribution.  The numerical solutions of the equation are in good
agreement with the simulations of the model.  We also find that the
behavior of the model is not sensitive to various changes of the rules
for the dynamics.

The paper is organized as follows.  In Sec.~\ref{sec:noblock}, we
define the model, and study the model without blocking.  We form a
differential equation, and its solution is compared with the
simulations of the model.  In Sec.~\ref{sec:block}, we study the
steady state of the full model with blocking, and compare it with DP.
We also study the evolution of the distribution of trapped particles.
In Sec.~\ref{sec:stability}, a few modification of the rules are
introduced to check the stability of the behavior of the model. A
brief summary of the results and limitations of the model are given in
Sec.~\ref{sec:conclu}.

\section{Model without Blocking}
\label{sec:noblock}

\subsection{Definition of the Model}

Consider a square lattice, rotated by $45$ degrees to the flow axis,
of width $W$ and length $L$, which is an idealized network model of
the filter pore space.  (Fig.~1).  The nodes and bonds of the lattice
represent pores and channels, respectively.  A periodic boundary
condition is applied in the transverse direction. Fluid containing
suspended particles is injected on the left side of the filter ($x =
1$ line), and exits the right side ($x = L$ line).  Suspended
particles, if not trapped, move along the local direction of the fluid
flow.  We consider particles of radius $R$, and assign channel radius
$r_j$ to bond $j$, where the radius is drawn from some distribution
$C(r)$. A bond with $r_j > R$ is called as a $B$ (big) bond, and other
bonds ($r_j \le R$) are $S$ (small) bonds.  Particles can move through
a $B$ bond without difficulty, while they would stuck in a $S$ bond.
Let the fraction of $S$ bonds be $p$.

The rules for the movement of a particle are defined as follows. A
particle is inserted at a randomly chosen node at the left end.  We
require that the particle always tries to move to right, the direction
of the fluid flow.  At a node, the particle has to choose a bond out
of the two bonds on its right for a movement.  We first consider the
case that no particle is trapped in the bonds.  Here, the particle
randomly chooses a bond with equal probability.  If the chosen bond is
a $B$ bond, the particle moves through the bond to the next node.  If
it is a $S$ bond, the particle is trapped in the bond.  The movement
of the particle is repeated until either the particle is trapped or
comes out of the filter.  We then insert another particle to the
filter, and the whole process is repeated.

We still have to define the rules involving bonds which contain a
trapped particle.  A reasonable rule is that if a particle is trapped
inside a bond, the entrance to the bond is blocked by the particle.
Note that only $S$ bonds can trap a particle.  There are two
possibilities involving blocked bonds.  If only one of the two bonds
for the movement is blocked, we always move the particle to the bond
which is not blocked.  If both bonds are blocked, the situation
becomes a little complicated.  The particle can not continue to move,
when it reaches such a dead end.  We solve the problem by blocking all
the paths leading {\em only} to dead ends.  The details of the
procedure will be discussed in Sec.~\ref{sec:block}.

The system with the rules defined so far is the main model, and will
be studied extensively for the most of the paper.  In this section,
however, we want to start with a simpler case of no blocking.  In this
case, a particle can move through a $S$ bond with a trapped particle,
and it will not be trapped.  In a sense, a $S$ bond with a trapped
particle is treated like a $B$ bond.  Thus, the effect of blocking by
a trapped particle is ignored.  The model without blocking is not
realistic, and its behavior is very different from that with blocking.
However, ignoring the effect of blocking makes the analysis of the
model tractable, and the method developed here will later be extended
to the full model with blocking.

\subsection{Simulation of the Model}

We present results of numerical simulations of the model without
blocking. The primary quantity of interest is the density field $\rho
(x,t)$ of trapped particles.  Here, $\rho (x,t)dx$ is defined as the
number of trapped particles in $[x, x+dx]$ divided by $2W$.  Also, $t$
is the total number of injected particles, which can be used as a
time-like variable.  In Fig.~2(a), we show $\rho (x,t)$ for several
values of $t$ and $p$.  For small $t$, the density field forms a
characteristic shape---two flat regions joined by a transition region.
For large $t$, the field seems to translate without much change of
shape.  If the shape of the field remains constant, one can show that
the curve should move with constant velocity $v(p) = 1 / 2Wp$.  The
velocity is defined as the amount of translation per injected
particle.  We translate the fields of Fig.~2(a) according to the
velocity as shown in Fig.~2(b).  The fields for different $t$ all seem
to collapse in a narrow region.  A closer inspection shows, however,
the shape changes slowly, but systematically, with $t$.  The width of
the transition region slowly increases with $t$.

A further point is that the shapes of $\rho (x,t)$ for different
values of $p$ seem to look the same.  One can roughly scale these
curves to a single curve as shown in Fig.~2(c).  The curves agree very
well with each other for small $t$.  Then, the width of the transition
region grows faster for larger values of $p$, and systematic
deviations from the collapse are visible as $t$ increases.  In scaling
the curve, we scale the width of the transition region by $1/p$.  The
argument for the choice is follows.  For a given $p$, the penetration
depth of a particle is order of $1/p$.  The width of the transition
region, which is the fluctuation of the penetration depth, is expected
to be similar to the penetration depth.

In the next section, we show that the simulational results discussed
here can be understood in terms of a mean field differential equation
for the evolution of $\rho(x,t)$.

\subsection{Evolution Equation of the Density Field}

We derive an approximate equation for the evolution of the density
field $\rho (x,t)$.  Consider the motion of a particle injected into
the filter.  Whenever the particle moves, there is certain probability
that the particle is trapped.  The average fraction of unoccupied $S$
bonds, which act as traps, at position $x$ is $p - \rho (x,t)$.  We
assume that the probability that the particle is trapped, while moving
from $x$ to $x+dx$, is $[p - \rho (x,t)] dx$.  Here, an approximation
is made which ignores the variation of the density field in the
transverse ($y$) direction.  Let $P(x,t)$ be the probability that the
particle arrives at position $x$ without being trapped.  Under the
above assumption, the probability of the particle to be trapped in
$[x, x+dx]$ is $[p - \rho (x,t)] P(x,t) dx$.  We thus arrive at
\begin{equation}
{\partial \over \partial x} P(x,t) = - [p - \rho (x,t)] P(x,t),
\end{equation} 
whose solution with condition $P(0,t) = 1$ is
\begin{equation}
P(x,t) = \exp[-\int_0^x p - \rho (x',t) dx'].
\end{equation} 
Since the density field increases by $1/2W$ for every new particle
trapped, the evolution equation for the density is
\begin{eqnarray}
\label{eq:tdrho}
{\partial \over \partial t} \rho (x,t) 
& = & {1 \over 2W} ~ [p - \rho (x,t)] ~P(x,t) \nonumber \\
& = & {1 \over 2W} ~ [p - \rho (x,t)] \exp[-\int_0^x p - \rho (x',t) dx'].
\end{eqnarray}
The derivation of (\ref{eq:tdrho}) involves another approximation.
The change of the density field per injected particle is assumed to be
proportional to the trapping probability which is {\em averaged} over
all possible trapping sites (mean field approximation).  On the other
hand, the relevant density field, and the one we have considered, is
obtained by injecting certain number particles (e.g.  $10,000$)
without taking an average after each injection.  The average is taken
only {\em after the whole injection} of particles.  The two procedures
are, in general, not equal.  The validity of the approximations in
deriving (\ref{eq:tdrho}) will be checked with simulations of the
model.

The evolution equation (\ref{eq:tdrho}) contains an integral in the
exponent, which makes further analysis less convenient.  The integral
can be eliminated by simple manipulations. Motived by the wave-like
behavior of the density field found in the simulations, we search for
a traveling wave solution---$\rho (x,t) = f[x - v(p)t]$.  Inserting it
into (\ref{eq:tdrho}),
\begin{eqnarray}
{\partial \over \partial t} f(x-vt) 
& = & -v {\partial \over \partial x} f(x-vt) \nonumber \\ 
& = & {1 \over 2W} ~ [p - f(x-vt)] ~ \exp[-\int_0^x p - f(x'-vt)dx'].
\end{eqnarray}
Differentiating the equation with respect to $x$, and after a little
rearrangement,
\begin{equation}
\label{eq:nldif}
{\partial^2 \over \partial x^2} f = -{1 \over p - f} ({\partial \over 
\partial x} f)^2 - (p - f) {\partial \over \partial x} f,
\end{equation}
which is a nonlinear differential equation.  

We can not obtain the analytic solution of the equation, and we
numerically solve it using a Runge-Kutta routine (d02haf) in the NAG
library.  The solution is calculated in interval $[0, 10]$.  We choose
$f = p$ at $x = 0$.  The boundary condition at the other end is a bit
subtle.  We choose $f$ to be close to, but not equal to, $0$ at $x =
10$.  Note that $f$ can not be $0$ due to the non-zero probability
that a particle passes through the interval.  We have tried $f =
10^{-3}, 10^{-4}, 10^{-5}$ with no essential difference in the result.
The value of $f$ is chosen to reveal the whole shape of the field in
the interval.  The value of $f$ at $x = 10$ changes the amount by
which the curve is translated, not the shape of the curve.  In
Fig.~3(a), we show the numerical solutions of the equation for several
values of $p$.  The shape of the field is very similar to that in
Fig.~2.  Furthermore, the solutions satisfy the same scaling as the
simulations.  Here, the scaling is almost perfect without any visible
deviation.  We also show both the numerical solution and the density
fields obtained by the simulations in Fig.~3(b).  There is good
agreement, especially at early $t$ of the simulation.  However, the
width of the transition region of the field from the simulations
gradually increases, as $t$ increases.  The equation (\ref{eq:tdrho})
seems to provide a good overall description of the results of the
simulations except the broadening of the interfaces.

What is the possible origin of the broadening?  Think of the filter
bed as a set of $W$ columns perturbed by the transverse coupling
between them.  The average number of particles injected in a column is
$n = t / W$.  The fluctuation of $n$ should be order of $\sqrt{n}$.
Consider a column in the filter.  Since the average position of the
transition region in the column is $\bar{x} = n / 2p$, the fluctuation
of $\bar{x}$ is $\delta x = \sqrt{\bar{x} / 2p}$.  Therefore, the
width of the transition region of the {\em whole} filter is affected
not only by the width $\omega_s$ of the transition region of single
column, but also by $\delta x$.  A rough estimate is that the
resulting width $\omega$ becomes $\sqrt{\omega_s^2 + (\delta x)^2}$.
which implies that the ratio $\omega / \omega_s$ is $\sqrt{1 + 2 p
\bar{x}}$.  Thus, the effective width increases with $t$ ($\bar{x}$), 
and the rate of the increase is larger for larger $p$, which are
consistent with the results of the simulations (Fig.~2).

\section{Full Model with Blocking}
\label{sec:block}

\subsection{Steady State Behavior}

Having obtained reasonable understanding of the model without
blocking, we proceed to a more interesting case where blocking is
present.  We first discuss the steady state of the model.  It will
later become clear that the information of the steady state plays a
crucial role in describing the evolution of $\rho (x,t)$.  A steady
state is reached when there are no more empty $S$ bonds which can be
reached by an injected particle, thus the density field $\rho (x,t)$
will remain constant.  Let the steady state density field be $\rho
_s(x)$.  We consider the steady state in the limit of the infinite
system size. In the steady state, if $p < p_c$, all injected particles
pass through the filter without being trapped.  On the other hand, if
$p > p_c$, all the paths leading to an exit from the filter are
blocked.  Here, $p_c$ is a threshold.  We start to see the similarity
of the present model to directed bond percolation (DP) \cite{s85}.  By
comparing the rules of the present model with those of DP on square
lattice, one can notice that the positions of trapped particles in the
steady state are identical to those of the blocked bonds connected to
a cluster in DP.  We thus expect the steady state density field $\rho
_s(x)$ to be described by DP.

We briefly review predictions of DP.  First, there is a percolation
threshold, $p_c$.  The exact value of $p_c$ is not known, and the best
estimate for bond percolation on square lattice is $0.355299(1)$
\cite{egd88}.  Note that $p$ represents the blocking probability, not
the conducting probability commonly used for percolation.  We discuss
the behavior in three separate regimes.

{\bf $p = p_c$:} There exists a spanning cluster of unblocked bonds.
The mass of a spanning cluster can be calculated as follows.  The
probability that a bond belongs to a spanning cluster $P_{\infty}(p)$
scales as $|p - p_c|^{\beta}$.  Since the correlation length in the
longitudinal direction $\xi_{\parallel}$ scales as $|p -
p_c|^{-\nu_{\parallel}}$,
\begin{equation}
P_{\infty}(p) \sim \xi_{\parallel}^{-\beta / \nu_{\parallel}}.
\end{equation}
The total mass of a spanning cluster is $P_{\infty}(p)$ times the
width $2W$ and the length $\xi_{\parallel}$ of the cluster---$2 W
\xi_{\parallel}^{1 - \beta / \nu_{\parallel}}$.  Then, the mass of 
a spanning cluster in $[x,x+dx]$ divided by $2 W$, which behaves 
the same as $\rho _s(x)dx$, is
\begin{equation}
\label{eq:rhopc}
\rho _s(x)dx \sim x^{-\beta / \nu_{\parallel}}dx.
\end{equation}
Using the best estimates for $\nu_{\parallel}$ and $\beta$
($1.7334(10)$ and $0.277(1)$ respectively), $\beta / \nu_{\parallel}$
is determined to be $0.1598$ \cite{egd88}.  The steady state density
field at the threshold $p_c$ decays as a power law.

{\bf $p < p_c$:} There also exists a spanning cluster.  Following the
formalism in DP, we propose a scaling ansatz
\begin{equation}
\label{eq:rholpc}
\rho _s(x) \sim |p - p_c|^{\beta} ~g(x/\xi_{\parallel}),
\end{equation}
where $g(z)$ is a scaling function to be determined.  The scaling is,
as in DP, expected to be valid near $p_c$.  The function $g(z)$ has to
satisfy certain properties.  Consider the limit of $p \to p_c^{-}$,
which results in $\xi_{\parallel} \gg x$.  Since the density field has
to approach (\ref{eq:rhopc}), $g(z)$ should behave as $z^{-\beta /
\nu_{\parallel}}$ for $z \ll 1$.  On the other hand, the density field
has to approach $P_{\infty}$ as $z \gg 1$, which implies $g(z) \sim
1$.  In sum,
\begin{equation}
\label{eq:scaleg}
g(z) \sim \left\{ \begin{array}{ll}
                  z^{-\beta / \nu_{\parallel}} & \mbox{if z $\ll$ 1}, \\
                  1                            & \mbox{if z $\gg$ 1.}
                  \end{array}
\right.
\end{equation}

{\bf $p > p_c$:} There is no spanning cluster.  In the regime, we
propose a scaling ansatz
\begin{equation}
\label{eq:rhogpc}
\rho _s(x) \sim x^{-\beta / \nu_{\parallel}} ~h(x/ \xi_{\parallel}),
\end{equation}
where $h(z)$ is another scaling function.  If we take the limit of $p
\to p_c^{+}$, the density field has to approach (\ref{eq:rhopc}),
which requires $h(z) \sim 1$ as $z \ll 1$.  No new information about
$g(z)$ can be obtained in the other limit of $z \gg 1$.

We present results of the numerical simulations to compare with above
predictions.  In order to obtain a steady state, one can inject
particles one by one, until no particle can be trapped, literally
following the definition.  The procedure is quite time-consuming, and
there is a much faster way to determine the steady state density
field.  The method is based on the ``burning'' algorithm originally
used to study the properties of a percolation cluster \cite{hhs84}.
The state state density field is determined in a single ``sweep'' of
the system.  The method will be discussed in Appendix A.  In the
insets of Fig.~4(a) and (b), we show $\rho _s(x)$ for several values
of $p$, determined by the method.  The density field at $p_c$ decays
as a power law with an exponent consistent with (\ref{eq:rhopc}).  We
then scale these $\rho _s(x)$ according to the predictions of
DP---(\ref{eq:rholpc}) and (\ref{eq:rhogpc}).  All the curves seem to
collapse well into two curves, one for $p < p_c$ (Fig.~4(a)) and the
other for $p > p_c$ (Fig.~4(b)).  Only small deviations can be seen
for the values of $p$ away from $p_c$.  Note that all the parameters
used for the scaling (e.g., $p_c, \beta$) are those of DP, and no free
parameters are used.  The scaled curves, which are the scaling
functions $g(z)$ and $h(z)$, also satisfy the properties discussed
before.  The scaled curve of Fig.~4(a), which is $g(z)$, decays as a
power law for small $z$, and approaches a constant for large $z$.
Also, the curve $h(z)$ in Fig.~4(b) approaches a constant for small
$z$, and seems to decays as an exponential.  Thus, the comparison with
the numerical simulations confirms that the steady state density field
is well described by DP.

\subsection{Evolution of the Density Field}

We discuss the evolution of the density field for the full model with
blocking.  The simulation of the model poses a subtle problem.
Consider a particle moving in the filter.  If both of the bonds
available to the particle are blocked, the particle can not continue
to move.  What should be an appropriate rule for the movement?  In a
real situation, a particle chooses a channel according to the amount
of fluid flow in the channel.  Since the fluid flow in the channels
leading {\em only} to dead ends will be very small, particles rarely
go to these channels.  In the present simulation, we remove all the
paths leading {\em only} to blocked bonds.  To identify such a path,
one has to consider more than a local geometry, since all the paths
connected to the bond have to be traced.  We have developed a method
based on the burning algorithm \cite{hhs84}.  The method is similar to
the one used to remove the ``dangling'' bonds of an infinite
percolation cluster.  The detailed description of the method will be
given in Appendix B.  In Fig.~5, we show the evolution of the density
field for several values of $p$.  For small $p$, the overall shape of
the field is similar to the no blocking case (Fig.~2).  There are two
small differences, though.  The steady state value of the density
field for $x \gg 1$ is smaller than $p$, compared to the value of $p$
for the model without blocking.  The difference is due to the fact
that some of $S$ bonds are not accessible to injected particles.
Also, the width of the transition region for the model with blocking
is a bit larger.  As $p$ increases, even the overall shape of the
field becomes different from that without blocking.  The width of the
transition region becomes quite large (comparable to the length of the
system in some cases).  The density field for very small $t$ is
exponential, in agreement with the previous simulations
\cite{gahg96}.

We quantify the transition region by defining the average position
$\bar{x}$ and the width $\delta x$ of the region.  For $p < p_c$, the
inflection point of the field $\rho (x,t)$ is a suitable criterion for
$\bar{x}$.  We numerically calculate the spatial derivative $\rho ^{'}
(x,t) \equiv -\partial_x \rho (x,t)$ using the smoothed data of $\rho
(x,t)$.  The results are not sensitive to the exact procedure for the
smoothing.  The resulting field $\rho ^{'}(x,t)$ is a bell shaped
curve, where the position of the maximum is the inflection point.  We
define $\bar{x}$ and $\delta x$ as the mean $\langle x
\rangle _d$ and the standard deviation $\sqrt {\langle x^2 \rangle _d
- \langle x \rangle _d^2}$ of $\rho ^{'}(x,t)$, respectively.  Thus,
$\langle A \rangle_d$ is defined as
\begin{equation}
<A>_d = {\int_0^{L} \rho ^{'}(x,t) ~A(x,t) ~dx \over 
         \int_0^{L} \rho ^{'}(x,t) ~dx}.
\end{equation}
The average position $\bar{x}$ and the width $\delta x$ for several
values of $p$, where $p < p_c$, are shown in Fig.~6(a).  Comparing the
values with the density fields (Fig.~5) confirms that these values are
reasonable representations of the transition region.

Unfortunately, the above procedure can not be applied for $p > p_c$.
Here, the inflection point of the field, if it exists, is not a
reasonable representation of the mean position of the transition
region.  The density field behaves like a decaying exponential.  We
define $\bar{x}$ and $\delta x$ as the mean $\langle x \rangle_u$ and
the standard deviation $\sqrt{\langle x^2 \rangle _u - \langle x
\rangle ^2_u}$ of $\rho (x,t)$, respectively.  Thus, $<A>_u$ is
defined as
\begin{equation}
<A>_u = {\int_0^{L} \rho (x,t) ~A(x,t) ~dx \over 
         \int_0^{L} \rho (x,t) ~dx}.
\end{equation}
In Fig.~6(b), the values of $\bar{x}$ and $\delta x$ obtained
following the procedure are shown.  Again, the values seem to be
reasonable representations of the transition region.

How do we understand the evolution of the field?  Is there any
equation similar to (\ref{eq:tdrho}) which can be used for the
situation?  To answer the question, we inspect (\ref{eq:tdrho}) again.
In the equation, the trapping probability of a particle passing
through a channel is assumed to be $p - \rho (x,t)$, the fraction of
empty $S$ bonds.  In other words, we assume that {\em all} $S$ bonds
will eventually trap one particle.  One of the effects of blocking is,
however, to make some of $S$ bonds inaccessible to the injected
particle.  For the model with blocking, the fraction of accessible $S$
bonds is $\rho _s(x)$ instead of $p$.  It thus seems reasonable to
replace $p$ with $\rho _s(x)$ in (\ref{eq:tdrho}), when blocking is
allowed.  The proposed equation for the model with blocking is
\begin{equation}
\label{eq:tdrhoblk}
{d \over dt} \rho (x,t) = {1 \over 2W}  ~ [\rho _s(x) - \rho (x,t)] 
\exp[-\int_0^x \rho _s(x') - \rho (x',t) dx'].
\end{equation}
We numerically solve the equation, where we use the steady state
density field $\rho _s(x)$ measured in the previous section.  For
quantitative comparisons, we calculate the mean and the width of the
transition region of the resulting field using the methods discussed
before.  In Fig.~7, we show these quantities for several values of
$p$.  By comparing them with the ones obtained by the simulations
(Fig.~6), one can notice the overall behavior is essentially
identical.  Also, even their numerical values are in good agreements.
We thus believe that the modified equation (\ref{eq:tdrhoblk}) is a
good starting point for the description of the model with blocking.

\section{Stability of the Model}
\label{sec:stability}

In this section, we study a few variations of the model.  Our
objectives are twofold: we want the rules to be more realistic, and we
want to know how much the results (e.g., the density field) change
under the variations.  Only the quantities which are not sensitive to
the details of the model are meaningful, and can be compared with
experiments.

The first variation is to introduce the concept of flow induced
probability (FIP) \cite{s59,sds83,rf87}.  Consider a particle moving
through filter.  As the particle reaches a pore, it has to choose a
channel to continue its movement.  The exact rule for the choice is
complicated, and is not fully understood.  Still, it is a good
approximation to assume the particle chooses a channel proportional to
the amount of flow going through it, which is called flow induced
probability.  In the present model, the particle chooses a channel
with equal probability, if it is not blocked.  The problem in
introducing FIP to the model is that the flow field for the whole
system has to be calculated.  The calculation, not only is time
consuming, but also is against our intention of constructing a simple
model.

A simple solution for the problem can be obtained by noting the strong
correlation between the flow and the mobility of a channel.  It also
seems reasonable to assume that they are proportional to each other.
Thus, we implement FIP by assigning a mobility to channels, and assume
the amount of flow in a channel is proportional to its mobility.  The
mobility of a channel is determined as follows.  For a channel, one
chooses a radius $r$ drawn from distribution $C(r)$.  If one assumes
for simplicity that the channel length is on the order of the channel
radius, the mobility of the channel is proportional to $r^3$, where we
assume Poiseuille flow in a cylindrical tube.  How about distribution
$C(r)$ for the radius?  We first try a uniform distribution in the
interval $[1/2,1]$.  Thus, the probability that the radius is in
$[r,r+dr]$ is $2 dr$, if $1/2 < r < 1$, and is zero for other values
of $r$.  The new rule significantly changes local movements.  The
probability of choosing a channel can now differ by a factor of $5$ to
$6$.  In Fig.~8, we show the density field obtained by the numerical
simulations of the uniform distribution for several values of $p$.
There are small differences between the fields with and without FIP,
especially at small values of $t$.  The resulting difference is quite
small, considering the significant changes of particle movements.

We also study the field using $C(r)$ of the three dimensional random
close packing (RCP) of uniform spheres.  Here, we use the data for the
radial distribution of RCP in Ref.~\cite{m68}.  The exact procedure to
calculate $C(r)$ from the data is discussed in Appendix C.  In Fig.~8,
the resulting density field using the distribution of RCP is shown.
The field is again a little different from that without FIP, just like
the uniform distribution.  Flow induced probability does not
significantly changes the density field.  For later comparisons, we
also calculate the mean and the width of the transition region as
shown in Fig.~9.

The above implementation of flow induced probability can be included
in the framework of the evolution equation. The effect of FIP is that
it modifies the effective trapping probability of $S$ bonds.  For
given distribution $C(r)$, the probability that a particle to be
trapped $\Pi$ while passing through a channel is
\begin{equation}
\Pi = {\int_0^R \sigma(r) C(r) dr \over \int_0^\infty \sigma(r) 
C(r) dr},
\end{equation}
where $\sigma(r)$ is the mobility of a channel with radius $r$.  If
uniform conductance $\sigma(r) = \sigma_0$ is assumed, $\Pi$ returns
to the familiar value of $p$, the fraction of $S$ bonds.  The
effective distribution of $C(r)$ changes as particles are trapped in
$S$ bonds, making them inaccessible to further incoming particles.  We
model this change by removing blocked $S$ bonds out of the
distribution, and transferring them to $r > R$ part of the
distribution.  To be more precise, the effective distribution at given
$t$, $C(r,t)$, has to satisfy the following conditions.
\begin{equation}
\int_0^R C(r,t)dr = p - \rho (x,t),
\end{equation} 
and
\begin{equation}
\int_R^{\infty} C(r,t) dr = 1 - p + \rho (x,t).
\end{equation}
We choose to remove blocked bonds from $r < R$ part of the
distribution, thus $C(r,t) = [1 - \rho (x,t) / p] ~C(r)$.  We then
transfer the removed bonds to $r > R$ part of the distribution, thus
$C(r,t) = [1 + \rho (x,t) / (1 - p)] ~C(r)$.  In other words,
\begin{equation}
C(r,t) = \left\{\begin{array}{ll}
                (1 - \rho (x,t) / p) ~C(r) & \mbox{if $r \le R$}, \\
                (1 + \rho (x,t) / (1 - p)) ~C(r) & \mbox{if $r > R$.}
                \end{array}
\right.
\end{equation}
Thus, the probability to trap a particle passing through a channel
located at $x$ at time $t$ is
\begin{equation}
\Pi (x,t) = {(1 - \rho (x,t) / p) P_{L} \over
             (1 - \rho (x,t) / p) P_{L} + [1 + \rho (x,t) / (1 - p) ] P_{R}},
\end{equation}
where $P_L = \int_0^R \sigma(r) C(r)$ and $P_R = \int_R^{\infty}
\sigma(r) C(r)$.  The effect of blocking can be added by replacing $p$ 
with $\rho _s (x)$ in the above equation.  The resulting equation for
the evolution is
\begin{equation}
\label{eq:tdrhofip}
{d \over dt} \rho (x,t) = {1 \over 2W}  ~ \Pi (x,t) 
             \exp[-\int_0^x \Pi (x',t) dx'].
\end{equation}
We numerically solve the equation with blocking.  In order to compare
with the simulations with FIP, we calculate the mean and the width of
the transition region as shown in Fig.~10.  Comparing the solution
with the numerical simulations (Fig.~9), one notice good agreements
between them, which gives more confidence in the evolution equation.

Finally, we study the effect of the relaunching observed in experiment
\cite{gahg96}. There, when a particle passes near a trapped
particle, it occasionally kicks the trapped particle out of its site.
The kicked (or ``relaunched'') particle can be trapped again, or it
can move along the fluid flow.  The actual mechanism for the
relaunching, which is probably due to hydrodynamic interaction, is not
completely understood.  Here, we use a simple rule, which tries to
imitate the effect.  Consider a particle at a node.  If one of the
channels right to the node has a trapped particle, the trapped
particle will be kicked out of the bond with probability $q$.  Once
the particle is relaunched, it moves just like any other particle
\cite{gahg96}.  In Fig.~11, we show the density field obtained for
several values of $p$ and relaunching probability $q$.  In the
simulations, flow induced probability is included, and the
distribution $C(r)$ of RCP is used.  One can see the relaunching
changes a little the density field.  The density field does not seem
to be sensitive to the details of the rules.

\section{Conclusion}
\label{sec:conclu}

In this paper, we have studied a simple model for deep bed filtration.
The primary quantity of interest is the density field of trapped
particles.  The evolution of the density field is significantly
different depending on whether $p$ is below or above a threshold
$p_c$.  The density field and its evolution do not seem to depend on
the details of the rules.  In order to have some theoretical
understanding of the model, we have proposed a mean field equation for
the evolution.  The equation seems to describe well both qualitative
and quantitative behaviors of the model.

There are several things one should examine before taking the present
model seriously.  First of all, one has to check how sensitively the
results depend on the details of the rules.  The rules we have used
for particle movements---choosing a channel, the effect of blocking
and the relaunching of trapped particle---are too simple to be
realistic.  Thus, only behavior which is not sensitive to the rules
can be compared with experiments.  Here, we have studied a few
variations of the rules, and have found the behavior is not sensitive
to the changes, but more extensive study in this direction is
desirable.  The crucial next step to check the relevance of the model
to experiments, however, is to compare with network models
\cite{lk96}.  Network models are believed to be a faithful
representation of real porous media.  For example, the changes in the
flow field due to particle movements are taken into account in these
models.  If the simple model compares well with network models, it can
be used as a complimentary tool to study deep bed filtration, and in
particular the large scale behaviors of the system.

\section*{Acknowledgment}

We thank E. Guazzelli, C. Ghidaglia, L. de Arcangelis and S. Redner
for extensive discussions.  This work is supported by the Department
of Energy under grant DE-FG02-93-ER14327.

\section*{Appendix A Determination of the Steady State}

We describe the algorithm we use to find all accessible $S$ bonds.
Consider the square lattice shown in Fig.~1.  We assign variable
$c(x,y)$ to node $(x,y)$, and set its value to $0$. We start by
setting $c(x,y) = 1$ at all the nodes on $x = 1$ line.  We then check
the nodes on $x = 1$ line.  For site $(1,j)$, we check the two bonds
connected right to the node.  If the bond is a $S$ bond, we mark is as
$T$.  If the bond is $B$ bond, we set $c(x,y) = 1$ at the node
connected to $(1,j)$ node through the bond.  Having checked the nodes
on $x= 1$ line, we proceed to check the nodes along $x = 2$ line. If
$c(2,j) = 1$, we update the bonds and the nodes connected to $(2,j)$
node as described before.  We repeat the procedure until $x = L$ line.
After the update, the bonds marked as $T$ are the bonds which trap a
particle in the steady state.

\section*{Appendix B Removing the Dead Ends}

We describe the algorithm we use to remove all the paths which lead
{\em only} to dead ends.  The essential idea is to start from a dead
end, and trace back all the paths leading to it.  To be precise,
consider a network as shown in Fig.~1.  We start from the right end of
the filter, $x = L-1$ line.  We check all the nodes at the line.  For
node $(L-1,y)$, we check if both of the bonds right to the node is
blocked.  If both of them are blocked, we block the entrance to the
node.  In other words, we block any $B$ bond left to the node.  We do
not have to worry about $S$ bonds, since incoming particles will block
them.  After checking all the dead ends at $x = L - 1$ line, we go to
$x = L - 2$ line.  We check if both of the bonds right to node
$(L-2,y)$ are blocked, and block appropriate $B$ bonds if necessary.
We repeat the procedure until $x = 1$ line.  After the sweep, all the
paths leading only to dead ends are blocked.

\section*{Appendix C Calculation of $C(r)$ of RCP}

We describe the method we use to calculate the channel radius
distribution $C(r)$ for the three dimensional random close packing
(RCP) of monosize spheres.  In essence, we can calculate $C(r)$ from
the nearest neighbor distribution $N(r)$ of RCP, where $N(r) dr$ is
the number of neighbors whose center lies $[r,r+dr]$ away from the
center of a reference particle.  We use the data for $N(r)$ in Ref.
\cite{m68}.  We scan the figure of $N(r)$ to obtain a postscript bitmap 
image file.  Then, we read the coordinates of non-empty pixels from
the file.  After simple rescaling, $N(r)$ can be reconstructed from
the pixel coordinates.  From $N(r)$, we generate three lengths
$r_{12}, r_{23}$ and $r_{31}$.  Here, $r_{ij}$ is the center-to-center
distance between sphere $i$ and $j$.  We take the channel size as the
radius of the sphere which barely fits in the hole formed by the three
spheres.  The present methods ignores the correlation between the
neighbor distances (e.g., $r_{12}$ and $r_{23}$).  However, the
comparisons of polygons and polyhedrons generated by the present method
with those by actual measurements confirm that the method is an
excellent approximation \cite{m68}.  The channel radius distribution
obtained here agrees well with the one in Ref. \cite{m68}.

\newpage
\section*{Figure Captions}

\begin{description}

\item [Fig.~1:] A schematic view of the model filter: nodes and bonds 
in the square lattice represent pores and channels in filter,
respectively.  $S$ bonds are shown with thin lines, where thick lines
are used for $B$ bonds.  A moving particle is shown with an empty
circle, and a trapped particle with a filled circle.  The arrows next
to bonds show possible movement of a particle.  For example, the
particle can not move to the $S$ bond with a trapped particle.

\item [Fig.~2:] (a) The density field $\rho(x,t)$ for $p = 0.2$ and 
$1.0$ is shown for several values of $t$.  The difference of $t$
between the successive fields is $10^4 ~p$.  (b) The fields in (a),
translated by $v(p) t$, are shown.  (c) The scaled fields for $p =
0.1, 0.5$ and $1.0$ are shown.  The scaled fields collapse very well
for small $t$, but deviations from the collapse are apparent for large
$t$. Here, we use $W = 500$ and $L = 100$, and all the fields are
averaged over $100$ samples.

\item [Fig.~3:] (a) The solution of (\ref{eq:nldif}) for $p = 1, 5, 
10$ is shown.  The scaling used for the fields is the one used for the
simulational data (Fig.~2(c)).  The fields exhibit excellent scaling
behavior.  Here, the fields are translated by $\delta$ so that the
centers of the transition regions coincide.  (b) The solution for $p =
1$ is shown with the density fields from the simulations.  There are
good agreements at early $t$.

\item [Fig.~4:] The scaled steady state density fields $\rho_s(x)$ of the
model with blocking is shown for (a) $p < p_c$: $p = 0.3193, 0.3367,
0.3457, 0.3504$, and for (b) $p > p_c$: $p = 0.3602, 0.3649, 0.3739,
0.3913$.  The fields before scaling are shown in the insets.  Also,
the field for $p = p_c$ is shown in the inset of (b).  Here, $W =
100$, $L = 500$, and the density fields are averaged over $100$
samples.

\item [Fig.~5:] Evolution of the density field with blocking is shown
for (a) $p = 0.3193$, (b) $p = p_c$, and (c) $p = 0.3913$.  Here, $W =
100$, $L = 500$, and the fields are averaged over $100$ samples.  The
difference of $t$ between the successive fields is $3000$ for (a) and
(b), and is $1000$ for (c).

\item [Fig.~6:] The mean position and the width of the transition 
region from the simulation data like in Fig.~5: The width is shown (a)
for $p < p_c$: $p = 0.3193$, $0.3367$, $0.3457$, $0.3504$, and (b) for
$p > p_c$: $p = 0.3602$, $0.3649$, $0.3739$, $0.3913$.  The mean
positions are also shown in the insets.

\item [Fig.~7:] The mean position and the width of the transition 
region of the density field obtained from (\ref{eq:tdrhoblk}): the
width is shown (a) for $p < p_c$: $p = 0.3193$, $0.3367$, $0.3457$,
$0.3504$, and (b) for $p > p_c$: $p = 0.3602$, $0.3649$, $0.3739$,
$0.3913$.  The mean positions are also shown in the insets.

\item [Fig.~8:] The evolution of the density field with and without FIP 
for (a) $p = 0.3457$ and (b) $p = 0.3649$.  Here, $W = 100$ and $L =
500$, and the averages are taken over $100$ samples.  The difference
of $t$ between successive density fields is (a) $3000$ and (b) $1000$,
respectively.

\item [Fig.~9:] The mean position (inset) and  the width  of the 
transition region for the density fields of Fig.~8(a).

\item [Fig.~10:] The mean position (inset) and the width of the
transition region from the numerical solution of the evolution
equation (\ref{eq:tdrhofip}) are shown.

\item [Fig.~11:]  The evolution of the density field with relaunching 
is shown for several values of $q$ and for (a) $p = 0.1$ and (b) $p =
0.2$.  Here, $W = 100$, $L = 100$, and the field is averaged over
$100$ samples.  Here, we use $C(r)$ of RCP, and flow induced
probability is included.

\end{description}

\end{document}